\documentclass[twocolumn, a4paper, prx, longbibliography ]{revtex4-2}
\pdfoutput=1
\usepackage[utf8]{inputenc}
\usepackage[english]{babel}
\usepackage[T1]{fontenc}
\usepackage{amsmath, amsthm, amssymb}
\usepackage{enumitem}

\usepackage{physics} % for appendix math definitions

\usepackage{xcolor}
\definecolor{myblue}{RGB}{59, 143, 217}
\definecolor{myred}{RGB}{217, 61, 61}
\usepackage{hyperref}
\hypersetup{
	colorlinks=true,
	linkcolor=myred,
	filecolor=magenta,      
	urlcolor=myblue,citecolor=myblue}

\usepackage[capitalize]{cleveref}

\usepackage{tikz}
\usepackage{lipsum}

\usepackage{appendix}

\addto\captionsspanish{%
  
}

\newcommand{\reals}{\mathbb{R}}
\DeclareMathOperator*{\argmax}{arg\,max}
\DeclareMathOperator{\sign}{sign}

\newcommand{\bre}[1]{\left[ #1 \right]}
\newcommand{\ident}{\vb{I}}

\DeclareMathOperator{\expecOp}{\mathbb{E}}
\newcommand{\Expec}[2]{\expecOp_{#1}\bre{#2}}
\begin{document}

\title{Quantum circuit synthesis with diffusion models}

\author{Florian Fürrutter}
\affiliation{Institute for Theoretical Physics, University of Innsbruck, Technikerstr. 21a, A-6020 Innsbruck, Austria}

\author{Gorka Mu\~noz-Gil}
\email{gorka.munoz-gil@uibk.ac.at}
\affiliation{Institute for Theoretical Physics, University of Innsbruck, Technikerstr. 21a, A-6020 Innsbruck, Austria}

\author{Hans J.~Briegel}
\affiliation{Institute for Theoretical Physics, University of Innsbruck, Technikerstr. 21a, A-6020 Innsbruck, Austria}

\begin{abstract}
Quantum computing has recently emerged as a transformative technology.
Yet, its promised advantages rely on efficiently translating quantum operations into viable physical realizations.
In this work, we use generative machine learning models, specifically denoising diffusion models (DMs), to facilitate this transformation.
Leveraging text-conditioning, we steer the model to produce desired quantum operations within gate-based quantum circuits. 
Notably, DMs allow to sidestep during training the exponential overhead inherent in the classical simulation of quantum dynamics—a consistent bottleneck in preceding ML techniques. 
We demonstrate the model's capabilities across two tasks: entanglement generation and unitary compilation.
The model excels at generating new circuits and supports typical DM extensions such as masking and editing to, for instance, align the circuit generation to the constraints of the targeted quantum device.
Given their flexibility and generalization abilities, we envision DMs as pivotal in quantum circuit synthesis, enhancing both practical applications but also insights into theoretical quantum computation.
\end{abstract}
\maketitle

Quantum computing is considered as a groundbreaking technology, holding promise across various domains, from fundamental research in physics~\cite{feynman2018simulating} or chemistry~\cite{mcardle2020quantum}, to machine learning~\cite{cerezo2022challenges} and optimization~\cite{farhi2014quantum}. 
While the technological realization of quantum processors is advancing, the current noisy intermediate-scale quantum (NISQ) era~\cite{preskill2018quantum} presents challenges due to device limitations, making it essential to design efficient quantum circuits tailored to the available resources and gate sets, requiring further research and expertise in finding optimal gate ansätze and methods for circuit construction.

This task, commonly known as quantum circuit synthesis, involves the following: 
given a target quantum state that the quantum processor aims to produce as output, one must design a sequence of basic elements, called quantum gates, that generate that quantum state from a fiducial / standard input state.
Similarly, instead of producing a desired quantum state, the target could be to implement a quantum algorithm or more generally a unitary operation.
%given a specific quantum state, algorithm, or more generally, a unitary defining a quantum operation, the goal is to create a sequence using an allowed set of quantum operations that successfully achieves the task. 
In particular, researchers have lately demonstrated significant advancements in circuit synthesis by harnessing machine learning techniques, ranging from reinforcement learning to generative models, as for instance in quantum state preparation~\cite{arrazola2019machine, bolens2021reinforcement, melnikov2018active}, ansätze prediction~\cite{he2023gnn, shen2023prepare, Zhang_2021}, circuit optimization~\cite{fosel2021quantum, ostaszewski2021reinforcement} or unitary compilation~\cite{zhang2020topological, moro2021quantum, sarra2023discovering, preti2023hybrid}. 
In most of these applications, ML methods rely on minimizing a cost function that compares the output of the resulting quantum computation to the desired one. 
This metric is generally classically hard to simulate~\cite{khatri2019quantum}, hindering the trainability of the models. 
Additionally, these methods are typically trained on specific configurations defined by their training data and struggle to adapt to novel, uncharted scenarios.

\begin{figure*}
    \centering
    \includegraphics[width=\textwidth]{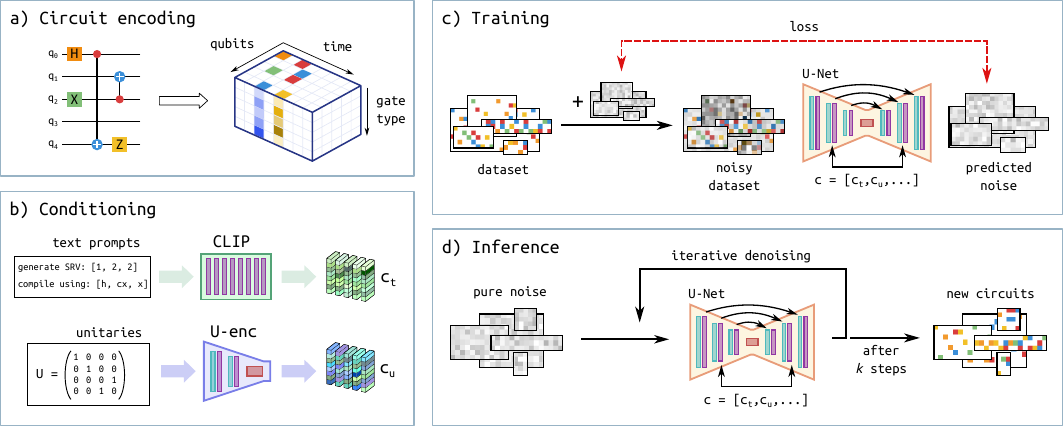}
    \caption{\textbf{Quantum circuit generation pipeline summary.} 
    \textbf{(a)} Quantum circuits are encoded in a three dimensional tensor, where each gate is encoded as a continuous vector of certain length (vertical direction), schematically represented here as a color (upper plane). 
    \textbf{(b)} Creation of the diffusion model's conditioning. Text is transformed into a continuous representation by means of a pre-trained CLIP encoder. In other cases, as for unitary compilation, an encoder is trained together with the DM to create the encoding of an input unitary.
    \textbf{(c-d)} Schematic representation of the training of the diffusion model and the posterior inference from the trained model. See text for details.}
    \label{fig:scheme}
\end{figure*}

To address these challenges, we use denoising diffusion models (DM)~\cite{sohl2015deep}, the current state-of-the-art machine learning generative method~\cite{rombach2022high}.  
Given a text prompt specifying a certain task, the model generates the desired quantum architecture. 
While we primarily focus on qubit-based quantum circuits, the model is straightforwardly extendable to other platforms such as measurement-based quantum computers~\cite{raussendorf2001one} or fermionic processors~\cite{gonzalez2023fermionic}. 
As any DM, the architecture is trained to \textit{denoise} corrupted samples from the training set, conditioned to the given text prompt. 
Hence, contrary to previous approaches, one avoids comparing the output of the generated quantum circuits to the target result. This ensures that no classical simulation of the circuit is needed and thus prevents the computation of intractable cost functions~\cite{khatri2019quantum}. 
Once trained, the method excels at generating correct, never seen circuits, allowing for instance the discovery of novel circuits decompositions, replacements, or compressions.
The architecture is highly flexible, allowing easy control of circuit characteristics such as length and number of qubits but also to account for the physical constraints of the target physical device, such as qubit connectivity. 
Additionally, in situations where generating training sets is costly, trained models can be updated with a handful of samples to acquire new skills.
We demonstrate the versatility of the method in two applications: 
First, we train the DM to generate circuits with varying degrees of entanglement, and use this task to benchmark the model’s capabilities;
Second, we train the DM to perform unitary compilation of unitaries with restricted sets of gates.

\section*{Circuit encoding, model and training pipeline}

In this section we present a comprehensive description of the proposed pipeline, namely, the encoding used to transform circuits into trainable tensors, the architecture of the denoising diffusion model (DM) and some key aspects of the model's training. All details necessary to reproduce the results presented here are available in the appendices and the accompanying code repository~\cite{github}, which also includes a series of examples that illustrate how diffusion models operate.

\paragraph*{Circuit encoding }
To train the DM, we take inspiration from~\cite{fosel2021quantum} and represent the circuits as three dimensional tensors. 
The first dimension encodes the qubit index, the second the \textit{time step} of gate placement (constraining to one gate per time step) and the third represents each gate (\cref{fig:scheme}a). 
We consider here a continuous gate representation of arbitrary length, fixed before training.
While in this work we assign random orthogonal high-dimensional representations to each gate, in more complex scenarios, for example optical devices in photonic setups~\cite{melnikov2018active}, such representations could be learned prior to the DM training.
To decode circuits from the tensor representation, we calculate the cosine similarity between the described gate and our existing gate set, selecting then the gate with the largest similarity.
Further details are given in Methods.

\paragraph*{Diffusion model}
DMs are a class of generative models aimed at learning a diffusion process that generates the probability distribution of a given training dataset~\cite{sohl2015deep}.
Practically, such process involves iteratively denoising an initially noisy sample to produce a high-quality one.
Led by the emergence of latent diffusion~\cite{rombach2022high}, and the ability to guide / condition their generation at will~\cite{ho2022classifier}, DMs have become the go-to method for generative applications in fields like image generation~\cite{podell2023sdxl}, audio synthesis~\cite{kong2020diffwave}, video generation~\cite{singer2022make}, and protein structure prediction~\cite{watson2023novo}.
In this work, we construct a pipeline similar to that of Ref.~\cite{rombach2022high}.
We first create the condition that will guide the generation of the DM (\cref{fig:scheme}b).
Text prompts are transformed into continuous representation by means of a pre-trained CLIP encoder, a neural network trained to associate visual content and textual descriptions, whose acronym relates to its training procedure (Contrastive Language-Image Pre-training)~\cite{radford2021learning}. 
Moreover, depending on the task at hand, further conditioning may be necessary. For instance, in unitary compilation, an encoded representation of unitary is appended to the encoded text representation. Here, the unitary encoder is trained alongside the DM (see Methods).

To train the DM itself, we corrupt the samples in the training set (i.e. the encoded circuits) with different levels of Gaussian noise (\cref{fig:scheme}c).
The model then learns to predict the noise present in each sample which, subtracted from the input samples, effectively denoises them.
We adapt the architecture from~\cite{rombach2022high} for the noise prediction model. The architecture is a typical U-Net consisting of an encoder-decoder layout with skip connections in between. We construct it by combining residual convolution layers and cross-attention layers~\cite{chen2021crossvit}. The latter implements the text conditioning naturally, as done by transformers in natural language processing~\cite{vaswani2017attention}. For our use-case, we propose some adjustments to the model to: 
i) account for the non-locality of qubit-connections.
ii) allow us to feed circuits of any size both in terms of the number of qubits and its length (see Methods). 
After training, one feeds a completely noisy tensor into the model which, guided by the chosen condition, iteratively denoises it, generating after some steps a high-quality sample (\cref{fig:scheme}d).

\section*{Results}

We now illustrate the capabilities of the method by applying it to two distinct problems: entanglement generation and unitary compilation. The former serves as a benchmark for showcasing the method's potential in different scenarios. The latter is a critical problem in the field, and will allow us to demonstrate the full capabilities of the method. Importantly, in both cases, the model is trained using the same denoising loss function, with different tasks achieved by modifying the training samples and their conditioning.

\begin{figure}
    \centering
    \includegraphics[width=\columnwidth]{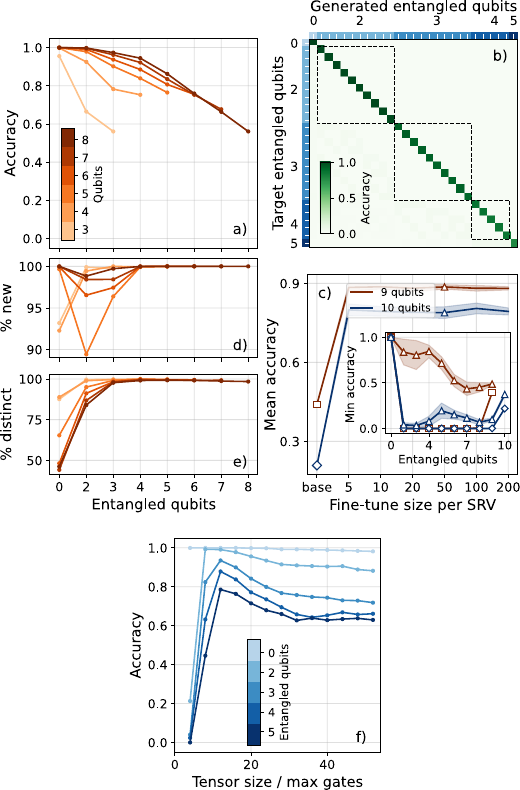}
    \caption{\textbf{Entanglement generation.} 
    \textbf{(a)} Model accuracy vs. the number of entangled qubits, for circuits of different qubits number.
    \textbf{(b)} Confusion matrix comparing, in each row and column, the input and generated SRVs for circuits of 5-qubits. For clarity, SRVs are grouped by their respective number of entangled qubits.
    \textbf{(c)} Mean accuracy over all SRVs vs. the number of samples used for fine-tuning the model in (a), for 9 and 10-qubit circuits. "\emph{base}" denotes the model's predictions before fine-tuning. Inset: minimum accuracy for the base (squares / rhombus) and fine-tuned model (triangles) on 50 circuits per SRV. Solid line and shaded area represent the mean and standard deviation over 10 fine-tune runs for both plots, respectively.
    \textbf{(d-e)} Percentage of new (not in training set) and distinct (not repeated in the generated sample) generated circuits.
    \textbf{(f)} Model accuracy vs. the input tensor size (which determines the maximum gate count), for 5-qubit circuits and different numbers of entangled qubits.
    Details on the data used in this figure given in Methods}
    \label{fig:srv}
\end{figure}

\paragraph*{Entanglement generation}
 
The automated design of experiments to generate new types of entanglement has been an exciting recent development in quantum physics~\cite{krenn2023artificial, melnikov2018active, PhysRevLett.116.090405}.
Here, our aim is to generate quantum circuits that produce states with specific entanglement, as described by their Schmidt rank vector (SRV), a numerical vector of dimension equal to the number of qubits that contains the rank of the reduced-density matrix for each subsystem~\cite{huber2013structure}. As we here deal with qubits, each element of the SRV is either 1 (non-entangled qubit) or 2 (entangled qubit).
To construct the training set, we randomly generate circuits made of $H$ and $CNOT$ gates with varying numbers of qubits (from 3 to 8) and gates (from 2 to 52). 
We then calculate their corresponding SRV, which can in this case take values $\in [1,2]$, and use it as text conditioning (see \cref{fig:scheme}).
We found that balancing the dataset both in terms of SRV but also w.r.t. to the circuit length greatly improved the training performance.
Once the model is trained, we generate new circuits for different SRV to assess its performance.  

We first observe that almost all continuous tensors generated by the DM can be faithfully transformed into circuits ($99.6$\% conversion rate).
This highlights the model's ability to learn gate encoding and proper gate placement within the circuit, ensuring that only one gate is placed at each time step.
In \cref{fig:srv}a, we present the accuracy of the DM, defined as the percentage of circuits generated from a given SRV-conditioning that indeed produce the desired SRV. 
To enhance readability, we average across SRV with the same number of entangled qubits, as the accuracy is usually constant within these.
\cref{fig:srv}b provides a detailed accuracy analysis for circuits of 5 qubits.
The majority of the circuits produce the correct entanglement, with the accuracy slightly dropping for larger numbers of entangled qubits.
This decrease arises because more entangled qubits demand a larger number of gates, thus presenting a more challenging problem. 
The variations in accuracy across different qubit numbers mainly reflect their representation in the training set.
Nevertheless, as sampling from the DM is a cheap operation, one strategy is to sample a few circuits with the desired property and select successful ones. While this may entail computationally costly operations, as one needs to simulate the quantum circuit and measure its properties, it needs to be computed only for the few circuits sampled.

In many cases, especially when increasing the number of qubits, creating a sufficiently large dataset to train the model becomes prohibitively expensive. In such scenarios, one can first train a model with simpler examples and then \emph{fine-tune} the model with few expensive ones. We demonstrate such feature by fine-tuning the model above to generate circuits of 9 and 10 qubits. As depicted in \cref{fig:srv}c, the base model can already generate with a reasonable average accuracy such larger circuits. However, a deeper examination shows that, while the method excels for certain SRVs, it completely fails in others (see e.g. the minimum accuracy in inset, square markers) . By fine-tuning the model with few circuits of 9 and 10 qubits per SRV (compared to the tens of thousands used for the original training), we greatly improve its accuracy, and most importantly banish the number of unattained target SRVs (inset, triangle markers).

Furthermore, the model exhibits an exceptional capability to discover new circuits that achieve the target SRV. 
\cref{fig:srv}d-e show that, across all SRVs and qubit sizes, the generated circuits are both new (not contained in the training set) and unique (not repeated within the generated set).
The model is hence able to generalize beyond the circuits given in the training set, providing a systematic and cost-effective method for discovering novel solutions to the task at hand.  
Lastly, the DM input tensor can be resized at will to account for different qubit numbers (as done above) and varying circuit lengths.
\cref{fig:srv}f shows that the accuracy of the model is almost constant no matter what the input tensor’s size is. The sudden drop for small sizes is due to the impossibility of generating entanglement with few gates.

\begin{figure}
    \centering
    \includegraphics[width=0.9\columnwidth]{ 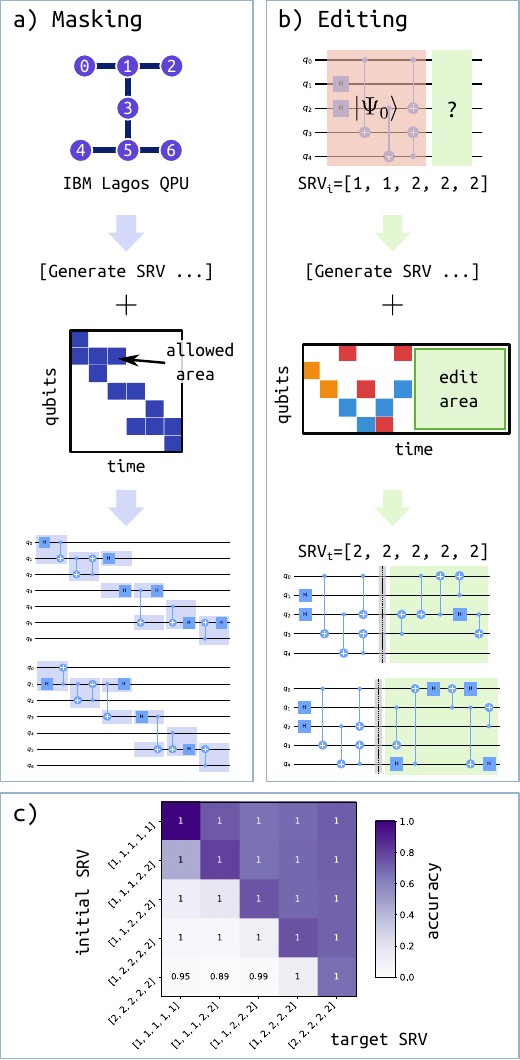}
    \caption{\textbf{Masking and editing circuits.}
    \textbf{(a)} Masking: the layout of a quantum processor can be embedded as a mask that prevents the model from placing gates at specific parts of the input tensor (white area).
    \textbf{(b)} Editing: parts of the circuit can be fixed to given gates prior to generation, for example to account for an initial input quantum state on which a desired quantum computation is to be performed.
    \textbf{(c)} Accuracy when editing circuits from an input SRV to a target SRV. Numbers highlight the fraction of initial circuits (256) where at least one solution was found within a sample of 1024 generated circuits.    
    }
    \label{fig:mask}
\end{figure}

\paragraph*{Masking and editing circuits}

In addition to resizing the input tensor, one can also fix parts of it, as proposed in Ref.~\cite{lugmayr2022repaint} for image editing.
This enables two operations: one prevents the model to place gates somewhere in the circuit, the other imposes the presence of certain gates.
Importantly, one does not require to retrain the model in any form, and the results presented in this section use the same model as employed above.

The first operation, \emph{masking}, offers a means to introduce specific structural constraints into the circuits. 
For instance, current quantum devices exhibit limited non-local qubit connectivity.
Performing operations between distant qubits, such as a two-qubit gate, must be broken into sequences of intermediate gates that swap the operation through the circuit~\cite{baumer2023efficient}, ultimately reducing gate fidelity.
However, such a constraint is not naturally represented in the circuit picture, where seemingly any qubit connectivity is feasible. 
As illustrated in \cref{fig:mask}a, to enforce the physical limitation of a known device onto the circuit, we mask desired parts of the model’s input tensor preventing any gate placement in the marked region. 
Then, this masked tensor together with the desired text prompt, is input into the model to generate the suitable circuit.
This approach also grants control over the number of qubits and circuit length e.g., by masking entire qubit rows, although the accuracy of the generated circuits  generally falls short compared to employing tensors of varying sizes, as done in the previous sections.

The second operation, \emph{editing}, allows us to fix specific gates throughout the circuit prior to the diffusion process.
From a practical perspective, we consider here that a quantum state of interest, in the form of a circuit, is given as an initial state in which further computations need to be performed (\cref{fig:mask}b).
For instance, starting with an input circuit of a certain SRV$_i$, we aim to transform it into a target SRV$_t$ (\cref{fig:mask}c).
We allow for this task circuits of maximum 20 gates, where at most the first 6 generate the initial SRV$_i$.
Notably, the model successfully handles tasks that involve increasing entanglement (elements above the diagonal), but encounters challenges when tasked with reducing entanglement (elements below the diagonal). 
This becomes particularly evident when a fully entangled state is given (lowest row). While on average the model produces accurate edits, for some circuits where all or most of the entanglement had to be destroyed, no correct editing was found.
We note that the latter is also the most challenging scenario, as disentangling highly entangled states requires the largest number of gates.
Moreover, our experiments (see Methods) suggest that the current model slightly correlates SRVs with circuit lengths, favoring shorter circuits for low entanglement. As editing towards such low entanglement is preceded by many gates for initial highly entangled states, the model has difficulties creating such target SRVs.
Nonetheless, in the worst case scenario the model is still capable of correctly editing 89.1\% of the input circuits.

\paragraph*{Unitary compilation}

\begin{figure}
    \centering
    \includegraphics[width=\columnwidth]{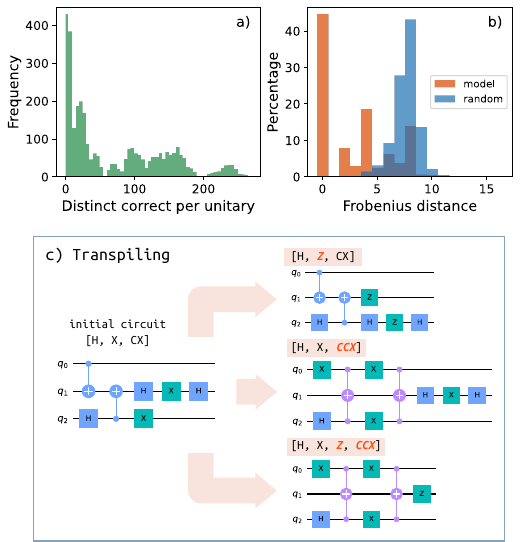}%\\
%    \vspace{0.4cm}
%    \includegraphics[width=0.7\columnwidth]{fig4c.pdf}
    \caption{\textbf{Unitary compilation.}
    \textbf{(a)} Number of exact and distinct circuit compilations, over 1024 generated circuits, for 3100 input unitaries.
    At least one exact circuit was found for 92.6\% of the unitaries.
    \textbf{(b)} Frobenius distance between the same 3100 target unitaries and circuits generated randomly (blue) or by the diffusion model (orange).
    \textbf{(c)} Given an input circuit, its variations can be produced by first calculating its unitary and then conditioning the model with this unitary plus different gate sets. 
    }
    \label{fig:compilation}
\end{figure}

We now train the model in one of the most crucial tasks in quantum circuit synthesis: compiling given unitaries into circuits, constrained to a certain set of gates.
To do so, we generate a variety of circuits with differing lengths and gates,  subsequently calculating their corresponding unitaries.
In previous tasks, generation was only conditioned to the information contained in the text prompt.
Here, we extend such conditioning by appending to the CLIP encoded text prompt, which now encodes the set of gates to be used, a learned encoding of the unitary (see \cref{fig:scheme}).
Just as before, the model is trained to denoise noisy input tensors, avoiding the exponential cost of computing distances between generated and target unitaries.
After training, new unitaries are fed into the model to evaluate its generative compilation capabilities.

We restrict here to 3 qubit circuits and unitaries computed from randomly generated circuits, ensuring that at least one exact implementation exists under the given constraints. Further, we only pick random unitaries that are not within the training set and new to the model.
For each input unitary, we generate 1024 circuits. Remarkably, the model successfully identifies the correct exact unitary for 92.6 \% of the 3100 tested unitaries.
In most instances the model finds multiple valid circuits for the same unitary (\cref{fig:compilation}a), allowing the prospective user to choose according to its needs. 
To gain further insights into the model's capabilities, we calculate the Frobenius norm based distance $\frac{1}{2}||U_t-U_g||_F^2$ between the target ($U_t$) and generated ($U_g$) unitaries (\cref{fig:compilation}b). 
As anticipated from the previous result, the majority of target unitaries can be compiled with norm zero.
When this is not achieved, the norm attained is smaller than what would be expected from a randomly generated circuit, suggesting that the model, while not capable of succeeding in the task, is heading in the right direction.
For those few failed cases, one can rely on further sampling or leverage the generated circuit as a starting point for other methods~\cite{weiden2023improving}.
In that sense, the masking and editing operations presented above can be performed in exactly equal manner for this task.

Interestingly, one can explore the compilation of the same unitary conditioned to multiple gate sets to uncover optimized solutions or identify unknown gate decompositions or compressions (\cref{fig:compilation}c).

\section*{Discussion}

In this work, we have demonstrated the potential use of diffusion models for quantum circuit synthesis.
As the development of quantum hardware progresses, there is an increasing need for efficient and versatile methods that convert desired quantum operations, as needed, e.g., in quantum algorithms~\cite{dalzell2023quantum} or quantum simulation~\cite{daley2022practical}, into practical physical implementations under given hardware constraints.
The approach introduced here addresses several persistent challenges in circuit synthesis: 
i)  through the model’s denoising loss, the method avoids the exponential overhead of classically computing cost functions comparing generated and target quantum circuits;
ii) the model’s flexibility enables adjusting text prompts and input tensors to produce circuits with specific lengths and topologies. This is especially useful for addressing physical device constraints or other boundary conditions and desiderata;
iii) Fine-tuning the model with a small number of samples from pre-trained models enables its training in situations where obtaining a full training set is prohibitively expensive.
We demonstrated these capabilities in two tasks: quantum state  preparation, specifically entanglement generation, and exact unitary compilation. The model not only tackles these tasks but also uncovers multiple novel solutions, offering further insights into the problem at hand.

The current work offers multiple extensions.
On the one hand, science greatly benefits from and needs interpretable machine learning solutions – models that not only provide predictions but also offer insights into how these solutions were found.
We believe that certain features of our proposed architecture can serve this interpretative purpose. 
For instance, as done in the context of image generations~\cite{niu2021review, wiegreffe2019attention, hertz2022prompt, li2023divide}, the U-Net's cross-attention maps can be used to reveal which parts of the circuit are deemed important by the machine for different text prompts.
On the other hand, perhaps the most direct benefit is the extension of the current method to continuous, parametrized gates, which can seamlessly be integrated into the current circuit encoding and training framework.
Moreover, the model can be extended to different platforms and applications, as e.g., the transpilation of quantum-gate-based operations to measurement-based quantum computations~\cite{raussendorf2001one, raussendorf2003measurement, briegel2009measurement}, discovering quantum experiments~\cite{melnikov2018active, krenn2021conceptual} or finding native compilations in fermionic~\cite{gonzalez2023fermionic} or anyonic~\cite{zhang2020topological} processors.
For each scenario, it is essential for the user to establish a suitable tensor encoding for its device's computational units, such as optical elements in a photonic processor or native gates in a fermionic processor. With a specific encoding set defined, a similar or even the same model presented here becomes readily adaptable to fulfill distinct user needs.

A key factor for the widespread application of the presented method is its scalability.
From a ML perspective, the diffusion model used here can be easily enlarged to reach the current state-of-the-art industrial models, given necessary computational power.
Indeed, a model on the scale of Stable Diffusion XL~\cite{podell2023sdxl} could potentially generate circuits of 1024 qubits and 1024 gates.
However, as in any circuit synthesis method, the main bottleneck arises from the quantum nature of the problem.
For instance, assembling a suitable dataset for such a large-scale model poses a considerable challenge, as it requires the classical computation of the unitary corresponding to each circuit or the SRV of its output.
Fine-tuning smaller, more manageable models for larger sizes, as demonstrated in \cref{fig:srv}c, may be practical for systems with tens  of qubits. 
Yet, alternative strategies may be necessary for handling hundreds or thousands of qubits.
Moreover, performing unitary compilation involves conditioning the model with a unitary matrix which grows exponentially with the number of qubits. Future iterations of the model here presented should explore more efficient conditionings, for example, based on Hamiltonians generating the unitary evolution of the system.
%We note nonetheless that these scalability challenges are shared with most circuit synthesis approaches and hence: 1) further development is needed in this field; 2) advancements in specific areas could greatly benefit the broader community, thus having a significant impact.

%Current experiments on training multi-task models (e.g. combining entanglement generation and unitary compilation) suggest that a large scaled training with an even bigger model and dataset could result in further emergent behavior (as done recently in image generation with Stable Diffusion XL~\cite{podell2023sdxl}). Leveraging such large scale training, the resulting model could find connections between different tasks and provide answers to more complex questions/prompts outside its training domain.
    
%TC:ignore

\paragraph*{Acknowledgements}
G.M.-G. acknowledges funding from the European Union.
H.J.B. acknowledges funding from the Austrian Science Fund (FWF) through [10.55776/F71] (BeyondC), the Volkswagen Foundation (Az: 97721), and the European Union (ERC Advanced Grant, QuantAI, No. 101055129).
Views and opinions expressed are, however, those of the author(s) only and do not necessarily reflect those of the European Union, European Commission, European Climate, Infrastructure and Environment Executive Agency (CINEA), nor any other granting authority.

\paragraph*{Code availability}
All the resources necessary to reproduce the results in this paper are accessible in Ref.~\cite{github}. The code is given in the form of a Python library, \verb|genQC|, which allows the user to train new models, generate circuits from pre-trained models as well as fine-tune the latter at will. The library also contains multiple examples that will guide the user through the various applications of the proposed method.

\paragraph*{Data availability}
The weights of the models trained for entanglement generation and for unitary compilation used to produce the figures presented in this manuscript can be found in Ref.~\cite{github}. The dataset are not shared due to space constraints but can easily be generated with the released code. All necessary details are provided in the different Methods sections.

\bibliography{biblio.bib}

%##################################################
\clearpage

\appendix

\makeatletter
\def\fnum@figure{Extended Data Figure \thefigure}
\makeatother
\setcounter{figure}{0}

\makeatletter
\def\fnum@table{Extended Data Table \thetable}
\makeatother
\setcounter{table}{0}

\begin{figure*}
	\centering
	\includegraphics[width=\textwidth]{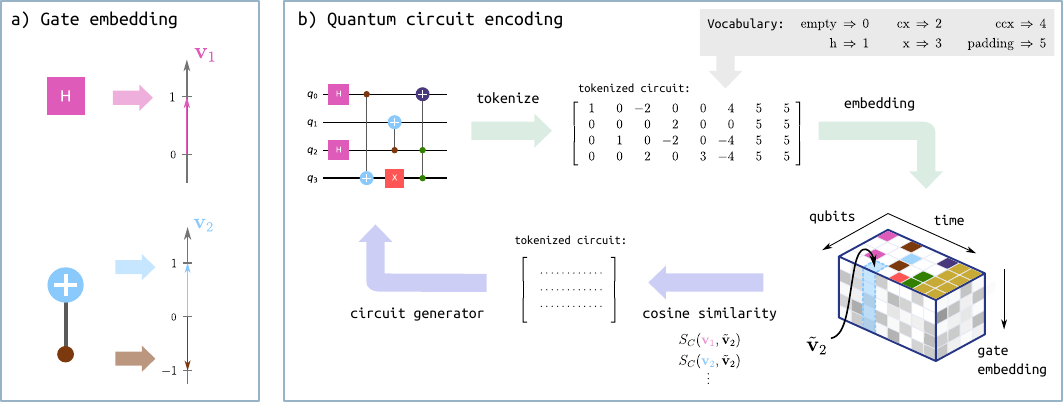}
	\caption{\textbf{Quantum circuit tensor encoding.} 
		\textbf{(a)} Schematic representation of the gate embeddings for a single and multi qubit gate.
		\textbf{(b)} Quantum circuit encoding and decoding pipeline. For encoding (green arrows), an input quantum circuit (top left) is first tokenized based on the proposed vocabulary. Then, the token matrix is transformed into a continuous tensor based on the chosen embeddings $\vb{v}_i$ (bottom right). In order to decode a continuous tensor into a circuit (blue arrows), we first use the cosine similarity between input embeddings and the ones assigned to existing tokens to generate a tokenized matrix, which is then transformed back into a circuit by means of the vocabulary. The transformation between circuits and tokens depends on such vocabulary, and can be changed at will to cope with the desired computing framework or platform. Further details are given in text.}
	\label{fig:app_tensEnc}
\end{figure*}

\section{Quantum circuit encoding} \label{se:app_tensorEnc}

In this section, we detail the procedure to encode quantum circuits into continuous tensors, which can then be fed into deep neural networks. Inspired by methods from the natural-language-processing (NLP) community, we first transform quantum gates into discrete tokens and then into continuous embeddings (see Extended Data Fig. 1). We will also refer to these embeddings as \emph{colors}, as done in \cref{fig:scheme}a.

The proposed encoding is constructed to account for:
\begin{enumerate}
	\item any sequence length (e.g., number of gates);
	\item any spatial size (e.g., number of qubits);
	\item any number of local or non-local spatial connections, as for instance two-qubit gates with target and control nodes;
	\item flexibility in the number of tokens, to possibly extend the number of gates at later stages of training or fine-tunings.
\end{enumerate}
\noindent
Points 1 and 2 are achieved by mapping the time (length) and space (qubits) dimensions each to one of the tensor's dimensions. Moreover, we require that only one gate can be present at each time step. Additional to these two axes, the tensor has a dedicated "gate dimension", of arbitrary size, used to describe each of the various gates. To address point 3, we describe the control and target nodes of a multi-qubit gate with the same numerical embedding but with opposite signs. As continuous vectors are used to describe gates, point 4 is seemingly addressed by assigning a new vector to each new gate. We note here that the proposed encoding is general and can be applied to any types of categorical time sequence. For instance, such continuous tensors can be used to represent parametrized gates or, extending beyond quantum circuits, optical devices in quantum optical experiments. 
\newline

\textbf{Encoding} -- The circuit encoding consists of two steps. First, we tokenize the circuit gates into an integer matrix representation. 
With it, each gate corresponds to an integer number, and we use a sign to specify the target and control nodes.
While one could work with such circuit representation, machine learning models' stability benefits from normalized continuous inputs, preferably centered around 0 and in the range [-1,1]. 
Hence, we transform each of the $N$ gates/tokens into a vector $\vb{v}\in\reals^d$, where the dimension $d$ is arbitrary. Again, the two nodes of a multi-qubit gate share the same $\vb{v}$ but with opposite signs.  We usually consider $d\geq N$, and choose the vectors $\vb{v}$ to be mutually orthogonal, defining a linearly independent set. The latter is not strictly necessary, but makes the decoding of model generated embeddings more robust to numeric variances. In this work, we consider $d=N+2$, where two tokens are used for padding and background. More on the former is presented in the Training section below.

As mentioned in the main text, in principle all the previous embeddings can be generated by trained machines (as done in NLP embeddings). Such training is usually performed prior to the diffusion model training. Importantly, if one trains it together with the denoising model, the trained embedder would learn to set all tokens to zero such that the denoising model can then extract the noise trivially.
\newline

\textbf{Decoding} -- The first step of the decoding entails transforming the continuous tensors $\vb{v}_{\mathrm{gen}}$ generated by the model back to tokens. For that, we choose the token $k$ whose embedding $\vb{v}_k$ has the highest absolute cosine similarity $S_C$ to $\vb{v}_{\mathrm{gen}}$:
\begin{equation} \label{eq:token_cos_sim}
	\tilde{k} = \argmax_{k}{\abs{S_C(\vb{v}_k, \vb{v}_{\mathrm{gen}})}}.
\end{equation}
To resolve the node type, we perform a second step:
\begin{equation}
	k = \tilde{k} \cdot \sign{S_C(\vb{v}_{\tilde{k}}, \vb{v}_{\mathrm{gen}})}
\end{equation}
Such calculation is performed for every space-time position of the tensor encoding, returning a tokenized integer matrix, which exactly describes a valid circuit. We refer to error circuits as those in which the model places two or more gates per time step or does not properly place either the target and control nodes of multi-qubit gates. Interestingly, we notice that the trained models matches the original embeddings with high precision (i.e.,$\vb{v}_{gen} \approx \vb{v}_k$) and assume one could "overload" the vector space $\reals^d$ with $N>d$ gates. This is further motivated by the results in visual image tasks \cite{rombach2022high}, where diffusion generated images are highly color accurate.

\section{Dataset} \label{se:app_dataset}
In this section we detail the procedure used to generate the datasets, with which both models presented in this work were trained (one for entanglement generation and one for unitary compilation). In both case we follow the same steps:
\begin{enumerate}[wide, labelwidth=!, labelindent=0pt]
	\item \textbf{Generate random circuits.} For each circuit, given the full set of gates considered for the current task, we first sample a subset of it. Next, we sample the number of gates to be placed from a uniform distribution with given lower and upper bounds. Then, we sample from the gate subset such a number of gates and place them in the circuit. Finally, we compute the condition associated with the generated circuits. For the entanglement generation task, we calculate the SRVs and convert them to prompts, e.g. "Generate SRV: [1, 2, 2]". For the unitary compilation task, we calculate the unitary and store it together with a prompt specifying the gate set that was used to sample the circuit, e.g. "Compile using: [cx, h, x]".
	\item \textbf{Optimize circuits.} We use the optimization-stage of the Qiskit \texttt{qiskit.compiler.transpile} function to optimize all the random circuits~\cite{Qiskit}. This step mainly removes duplicated consecutive gates and generally improves the quality of dataset. Training with such an optimized dataset indeed yields a significant increase of the model's accuracy. After optimizing, we delete all circuit duplicates.
	\item \textbf{Balance dataset.} We balance the dataset in a two-step procedure. First, we balance the text prompts, i.e., the SRVs or compilation gate sets, such that every prompt has the same number of circuits. In the second step, we balance the circuit lengths within the previously balanced text prompt bins. For that, we limit the number of circuits to a maximum threshold for each gate count. As the threshold we choose $q$-quantiles of the lengths within the text prompt bins. In particular, we chose $q=0.25$ for both the entanglement generation and unitary compilation task, and $q=0.05$ for the fine-tuning to 9 and 10 qubits in the entanglement generation task. Our experiments show that such balancing increases the dataset quality and improves the model performance.
\end{enumerate}
Using these steps, we produce datasets with the parameters specified in Extended Data Table 1. For the entanglement generation task, we use a multi-qubit dataset of 3 to 8 qubit circuits, with numbers of qubit and circuit length distribution presented in Extended Data Fig. 2. We trained the compilation task on a dataset with 2.5 million distinct 3-qubit circuits, created from 1.17 millions distinct unitaries.

\begin{table}[]
	\centering
	\begin{tabular}{l|ccc|p{19mm}}
		& qubits | & min gates | & max gates & gate pool  \\ \hline
		SRV	   & 3 & 2 & 16 & H, CX\\
		& 4 & 3 & 20 &\\
		& 5 & 4 & 28 &\\
		& 6 & 5 & 40 &\\
		& 7 & 6 & 52 &\\
		& 8 & 7 & 52 &\\ \hline
		
		SRV, fine-tune & 9  & 8 & 36 &  H, CX \\ 
		& 10 & 9 & 36 &   \\ \hline
		
		Compilation	       & 3 & 2 & 12 & H, CX, Z, X, CCX, SWAP \\
	\end{tabular}
	\caption{\textbf{Dataset sampling parameters.} Parameters used for the generation of the various training datasets used throughout the manuscript. SRV refers here to the entanglement generation task. We note that these parameters are chosen prior to the optimization step (step 2 in the Methods Dataset section), which effectively reduces the final gate number for some circuits.
	}
	\label{tab:app_dataset_sampling}
\end{table}

\begin{figure}
	\centering	
	\includegraphics[width=\columnwidth]{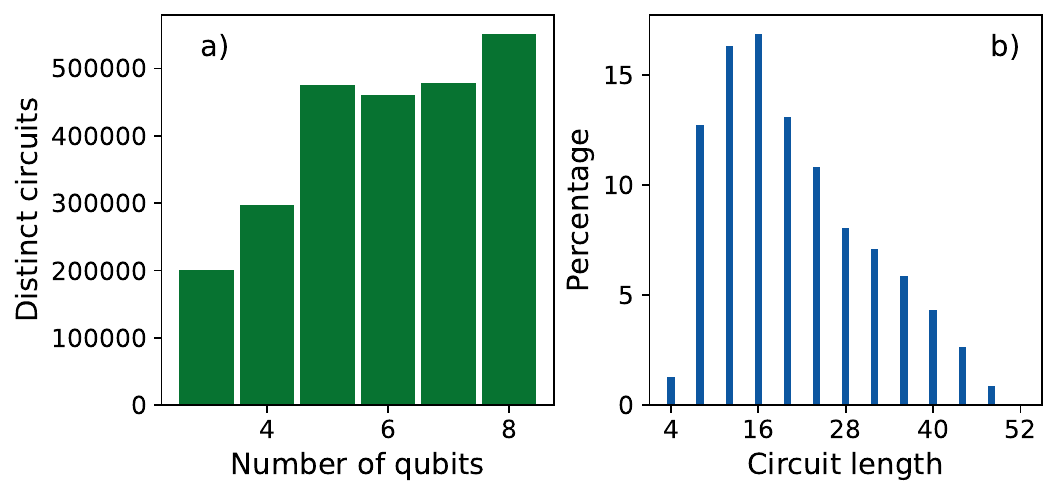}
	\caption{\textbf{Entanglement generation dataset distribution.} 
		Characteristics of the training dataset used for the entanglement generation task, sampled according to Extended Data Table 1 and balanced as described in Methods, Training section. Depending on the training step (max or bucket padding, see aforementioned section), we sample batches either from the whole dataset or buckets containing circuits of fixed number of qubits. 
		\textbf{(a)} Number of distinct circuits as a function of the number of qubits. For lower qubit counts, less distinct circuits exist, resulting in an inevitable lower number in the training dataset.
		\textbf{(b)} Distribution of circuit lengths, which are in this case multiples of the U-Net scaling factor 4, due to the length padding explained in Methods, Pipeline and Architectures section.}
	\label{fig:app_dataset_dist_srv}
\end{figure}

%#########################################################################
\section{Pipeline and architectures} \label{se:app_architectures}
In this section we present the details of the machine learning models, namely the denoising U-Net, the text encoder CLIP and the unitary encoder. Moreover, we explain how to combine conditions for the unitary compilation task. 
\newline

\textbf{U-Net} -- For the U-Net denoising model, we adjust the architecture of Stable Diffusion~\cite{rombach2022high}. An overview of the model is presented in~Extended Data Fig. 3a. The structure of the self and cross-attention layers is inspired by the Stable Diffusion model, which proposed a reordering of the normalization layers, compared to the original transformer architecture~\cite{vaswani2017attention}, and introduced the use of a residual connection. 

We design the architecture such that it accepts the circuit tensor encodings proposed in the Quantum circuit encoding section above. As discussed in the main text, to be useful for the proposed tasks, the model needs to account for: 
i) non-locality of qubit-connections.
ii) circuits of any size both in terms of the number of qubits and its length. 

To achieve this, all convolutions in the U-Net are performed \emph{only} on the time dimension. From a technical perspective, this entails having convolution kernels of size $1\times f$, where $f$ is the filter size in the time dimension, which varies throughout the architecture (see Extended Data Fig. 3). Such feature ensures that there is no locality bias induced in the qubit space dimension and, most importantly, that the perceptive field in this dimension is not restricted. This is crucial for instance in bigger circuits, where connections between outermost qubits exist. Computations over the spatial dimension are only performed in the self attention-layers which, most importantly, \emph{attend} simultaneously to the whole dimension, preventing any locality assumptions or size restrictions. To simplify the up and down scaling procedures contained in the U-Net, we restrict here to tensor encodings of lengths (time dimension) product of 4 (see Extended Data Fig. 2b). In practice we allow all circuit lengths and add padding tokens to the next higher multiple of 4.

Some conditions require an absolute position matching to the corresponding qubits (e.g. the elements of the SRVs strictly define which qubit has which Schmidt rank).  We account for this by adding an absolute 2D sinusoidal position encoding~\cite{vaswani2017attention} to the model input. To get a 2D position, we split the gate dimension in two chunks and apply on each a distinct encoding for the corresponding x and y coordinate. Inspired by~\cite{kazemnejad2023impact}, we tried different approaches: no positional encoding in either the spatial or time dimension, or introducing in both dimensions a causal attention mask. We noticed a significant accuracy drop when not using a 2D position encoding. We suspect that the application of the causal attention mask prevents the model seeing which gates it places in the subsequent time steps of the circuit. As any gate withing the circuit is able to alter the output state significantly, not having access to such information presents an insuperable challenge.

\textbf{Text prompt encoding} --
As described in the main text, we use a CLIP model~\cite{radford2021learning} with pre-trained weights from~\cite{Ilharco_OpenCLIP_2021} (version \textit{ViT-B-32} trained on \textit{laion2b\_s34b\_b79k}) to encode the text prompts. We discard the last layer and take the penultimate transformer output as encoding. The CLIP pipeline pads tokens to a sequence length of 77 elements. Since the last layer is discarded, the output-tensor has 512 channels i.e., a final shape of $77 \times 512$.

\textbf{Unitary encoder} -- 
Given a unitary matrix, we create an input tensor by splitting the unitary's real and imaginary parts into two channels, which is fed into the unitary encoder. Extended Data Fig. 3b schematically represents the used architecture. After an initial convolution layer, we introduce a 2D positional encoding layer, such as to encode the absolute position of the unitary entries. Then, we use a transformer encoder with self-attention layers, taking advantage of their global attention mechanism. Such global attention is key when inspecting a unitary matrix, as the information contained on it is highly non-local. We introduce in between the two attention blocks a down-scaling layer with kernel-size 2x2. The output tensor has size $l \times l \times 512$, where $l$ depends on the size of the input unitary.

\textbf{Conditioning} --
A key feature of the proposed model is its ability to generate quantum circuits based on a given conditioning. As usually done in this type of models, the input condition $\vb{c}$ is passed into the key and value inputs of the cross-attention layers of the U-Net. In our case, $\vb{c}$ can contain multiple information. In simple scenarios, as in entanglement generation, the CLIP encoded text prompt is the only condition. When performing unitary compilation, we concatenate the output of the CLIP encoder (size $77 \times 512$) with the flattened output of the unitary encoder (size $l^2 \times 512$), meaning a final conditioning of size $77+l^2 \times 512$. Last, the denoising time step $t$ (see below for details), encoded via a sinusoidal position encoding, is added to the residual part of the residual convolutions blocks inside the U-Net. 

\begin{figure*}
	\centering
	\includegraphics[width=\textwidth]{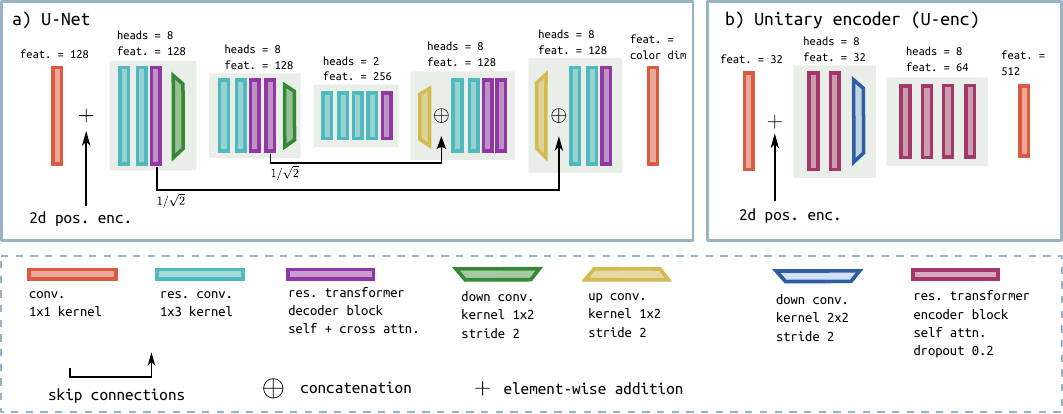}
	\caption{\textbf{Machine learning architectures.}
		\textbf{(a)} Scheme of the denoising U-Net architecture predicting the noise $\vb*{\epsilon}_\theta(\vb{x}_t, t, \vb{c})$. First, we project the input tensor features into a higher space through a convolutional layer (red) and then apply a 2D positional sinusoidal encoding. Then, we apply a typical encoder-decoder structure, with skip connections scaled with $1/\sqrt{2}$. The time step encoding $t$ is injected into residual  convolution layers (turquoise). The condition embeddings $\vb{c}$ are input to the residual transformer blocks (purple) as detailed in \cref{se:app_architectures}. All the transformer blocks have a residual connection. 
		\textbf{(b)} Scheme of the unitary encoder used to transform input unitaries into conditionings (see \cref{se:app_architectures})
	}
	\label{fig:app_architectures}
\end{figure*}

%#########################################################################
\section{Training} \label{se:app_Training}

In this section we present the details needed to reproduce the training of the presented models. Our models were trained using a single Nvidia RTX A6000 Ada graphical processing unit and an AMD Ryzen Threadripper Pro 5965wx processor. The training times vary between 12 to 24 hours depending on the particular application.

We train the DM according to the denoising diffusion probabilistic models (DDPM) procedure~\cite{DDPM_ho2020denoising}. We choose to parameterize the model as an $\epsilon$-predictor, i.e., at every time step $t$ the model learns to directly predict the noise $\vb*{\epsilon}_t$ of a noisy tensor $\vb{x}_t=\sqrt{\bar{\alpha}_t}\,\vb{x}_{0} + \sqrt{1-\bar{\alpha}_t} \,\vb*{\epsilon}_t$, where $\vb*{\epsilon}_t\sim \mathcal{N}(0, \ident)$, $\vb{x}_0$ is a sample of a training dataset with an arbitrary distribution $q(\vb{x}_0)$ and $\bar{\alpha}_t$ is a variance schedule. Here we set $\bar{\alpha}_t = \prod_{i=0}^{t}\alpha_i = \prod_{i=0}^{t} (1-\beta_i)$. Considering a denoising model with parameters $\theta$, whose noise prediction $\vb*{\epsilon}_\theta$ is conditioned on $\vb{c}$, its parameters are optimized to minimize the loss function
\begin{equation*}
	\mathcal{L} = \Expec{t\sim\mathcal{U}\bre{0, T},\, \vb{x}_0\sim q(\vb{x}_0), \,  \vb*{\epsilon}_t\sim \mathcal{N}(0, \ident)}{\norm{\vb*{\epsilon}_t - \vb*{\epsilon}_\theta(\vb{x}_t, t, \vb{c})}_2^2}.
\end{equation*}
In order to sample according to the given condition, we use classifier-free-guidance (CFG) \cite{ho2022classifier}. During training, we set the condition $\vb{c}$ to an empty token $\varnothing$ with a probability of 10\%, to train the model both for conditional and unconditional predictions. To reduce the exposure bias of the train-inference mismatch, we apply input perturbation \cite{IP_ning2023input} with $\gamma=0.1$. As variance schedule, we use the cosine beta schedule from \cite{betaSched_nichol2021improved}. We train for $T=1000$ diffusion steps and utilize the Adam optimizer~\cite{kingma2014adam} with one-cycle learning rate policy~\cite{smith2019super}.  
All further training parameters are listed in Extended Data Table 2.

For the compilation task, we simultaneously train the unitary encoder with the U-Net, i.e., treating the corresponding part of the condition as $\vb{c}_\Phi(U)$, where $U$ is a given unitary and $\Phi$ are the parameters of the unitary encoder. The CLIP model for the text prompt encoding is kept frozen. In our dataset, the majority of the unitaries have a single unique circuit implementation due to extensive random sampling in the unitary space.  This results in over-fitting, as the model can just memorize a one-to-one mapping of the unitary to a tensor encoding. We mitigate this by implementing dropout in the U-enc. Despite some overfitting during training, the model performs well when compiling new unitaries, as we show in~\cref{fig:compilation}. For the entanglement generation task, training the model shows a steady decrease in the validation loss, resulting from many circuits that correspond to the same "overloaded" SRV (see Method's Dataset section).

%------------------------------------------
A general quantum circuit synthesizer must be able to generate circuits with different sizes, i.e., different number of qubits and max number of gates. Translating this to image generation essentially means producing images with different sizes and aspect ratios. Tackling such problem is an ongoing research topic, which has recently been addressed by Stable Diffusion XL~\cite{podell2023sdxl}. Image generation datasets consist of many images with different sizes. Conventionally, all images are cropped, padded, resized or even discarded to one global size. Stable Diffusion XL introduces two procedures to mitigate this issue: micro-conditioning, i.e., conditioning on the original size of the images, and multi-aspect (ratio) fine-tuning.
In the presented model, we implement the latter in a two step process, as proposed in Ref.~\cite{podell2023sdxl}: 

\begin{enumerate}[wide, labelwidth=!, labelindent=0pt]
	\item \textbf{Max padding} -- First, we train on a single tensor size. For this, we calculate the maximum tensor size of the whole dataset (i.e., the most qubits and, separately, the highest number of gates). We then apply padding to all circuits smaller than this maximum size. In order to pad, we use an additional "padding" token and apply the same embedding method as for gates (as mentioned in the Quantum circuit encoding section and shown in Extended Data Fig. 1b). 
	
	\item \textbf{Bucket padding} -- After that, we split the dataset into \emph{buckets}, each containing circuits of a given qubit number. Then, the model is fine-tuned with batches from such buckets, ensuring that all circuits within a batch have the same number of qubits, but still allowing these to have varying lengths. Our experiments show a great improvement in the accuracy of the model when this last step was implemented.
	
\end{enumerate}  
Importantly, for both training steps and for each training batch, we cut the circuits to the longest circuit in the time dimension within the batch, removing unnecessary length padding and speeding up calculations. We believe that the model seeing the padding token for circuits smaller than the longest one within the batch is beneficial, as it can learn that the number of actual gates can be smaller than the tensor size.

\begin{table}
	\centering
	\begin{tabular}{l|c|c|c|r}
		model    & step & method      & lr                & epochs  \\ \hline
		SRV	3 to 8 qubits         & 1    & max pad     & $3\cdot 10^{-4}$ & 75      \\
		& 2    & bucket pad  & $5\cdot 10^{-5}$ & 50         \\ 		                    
		Fine-tune to 9 or 10 qubit	   &      & bucket pad  & $5\cdot 10^{-5}$ & 25         \\ \hline	
		Compilation 3 qubits       &      & bucket pad  & $3\cdot 10^{-4}$ & 150    \\ \hline
	\end{tabular}
	\caption{\textbf{Training parameters.} For the multi-qubit entanglement generation task (referred here as SRV), we use two training steps as explained in Methods, Training section. The fine-tuning and compilation trainings are done with circuits with fixed qubit numbers (either 9 and 10 for entanglement generation or 3 for unitary compilation). Hence, no special padding is needed in the space (qubit) dimension. Length padding is done per training batch. For all trainings we use a batch size of 256.}
	\label{tab:app_train_parameters}
\end{table}

\begin{table*}
	\centering
	\begin{tabular}{l|ccp{19mm}rccc|p{50mm}}
		Task                     & qubits | & max gates | & gate-pool | & samples | & denoise steps | & $t_\mathrm{start}$ | & $g$ &  Notes \\ \hline
		\cref{fig:srv}a, d, e    & 3 to 8  & 16        & H, CX 	& 8192		& 20		    & & 7.5 & samples per \# entangled qubits \\
		\cref{fig:srv}b          & 5       & 16 		& H, CX 	& 8192 		& 20 			& & 7.5 & samples per SRV                 \\
		
		\cref{fig:srv}c          & 9 to 10 & 16 		& H, CX 	& 512 	& 20		 	& & 7.5 & samples per \# entangled qubits, 10 fine-tune runs \\
		\cref{fig:srv}c inset          & 9 to 10 & 16 		& H, CX 	& 16384 	& 20		 	& & 7.5 & samples per \# entangled qubits, 10 fine-tune runs \\
		\cref{fig:srv}f          & 5       & 4 to 52 	& H, CX 	& 8192 		& 20 			& & 7.5 & samples per \# entangled qubits per max gates \\
		\hline
		
		\cref{fig:mask}a    	  & 7      	& 16  		& H, CX 	&     		& 40 			& 39 & 7.5 &  no initial gates  \\ 
		\cref{fig:mask}b         & 5       & 16 		& H, CX 	&     		& 40 			& 39 & 7.5 &  one initial circuit with 6 gates   \\ 
		\cref{fig:mask}c         & 5       & 20 		& H, CX 	& 1024 		& 40 			& 38 & 7.5 &  256 initial circuits with 5 to 6 gates\\ \hline
		
		\cref{fig:compilation}a, b  & 3 & 12 & H, CX, Z, X, CCX, SWAP & 1024 & 20 & & 7.5 & samples per unitary; 3100 unitaries with 2 to 12 gates \\ \hline              
	\end{tabular}   
	\caption{\textbf{Sampling parameters.} Parameters for all the results reported throughout this paper. Here $g$ is the CFG guidance-scale. For editing and masking $t_\mathrm{start}$ is the time step from which we start denoising (analogous to img2img \cite{img2img_meng2022sdedit}).}
	\label{tab:app_figure_parameters}
\end{table*}

\begin{figure}
	\centering
	\includegraphics[width=0.75\columnwidth]{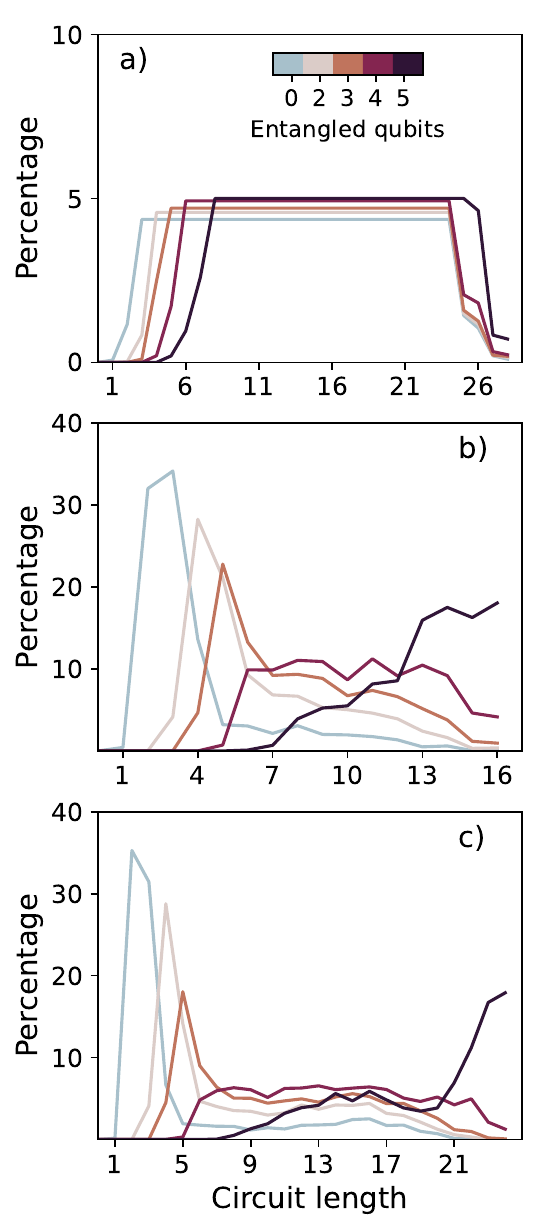}
	\caption{\textbf{Generated circuit lengths distributions.} Distribution of circuit lengths w.r.t. to the number of entangled qubits for: \textbf{(a)} the training (balanced) dataset of \cref{fig:app_dataset_dist_srv} filtered for 5 qubit circuits; \textbf{(b-c)} generated circuits with an input tensor constraining a maximum of 16 and 24 gates, respectively.
	}
	\label{fig:app_model_gen_len_distribution}
\end{figure}

%------------------------------------------
\subsection{Fine tuning}

In \cref{fig:srv}c we showed that a trained model can be fine-tuned after training and learn to solve new tasks. In that case, we showcased how it is possible to improve the accuracy of the model in bigger qubit counts: from the original 3 to 8 qubit model to 9 and 10 qubits. To do so, we perform an extra step of the bucket padding step, with circuits of 9 and 10 qubits, with parameters given in~Extended Data Table 2. In this case, we consider each qubit count as a completely different training. It is expected that, due to the fine-tuning, the accuracy of the model on the original task (e.g., 3 to 8 qubit circuits) may decrease. Here, we focused only on improving the accuracy of the new tasks. If one wants to keep the performance of old tasks, a mixed-qubit training can be used, where the new expensive circuits are combined with a dataset of already know tasks, e.g., the latter can be automatically generated by the model beforehand, as proposed by DreamBooth~\cite{DreamBooth_ruiz2023dreambooth}.

\section{Inference and benchmark}
Once trained, we sample (infer) new circuits from the model in a variety of scenarios. To sample a new circuit we provide the model with condition $\vb{c}$ and a noise input tensor $\tilde{\vb{x}}_T\sim\mathcal{N}(0, \vb{I})$. The noise tensor imposes the size of the output circuits, i.e., the number of qubits and the maximal number of gates the model is able to place. As the model has seen a variety of padded circuits during training, it is able to place the padding token if needed. Hence, a good strategy is to input sufficiently long tensors to the machine and allow the model to restrict the total number of gates by placing the padding token. We believe such feature underlies the stability of the accuracy over increasing tensor sizes shown in \cref{fig:srv}f.

As sampling method, we use the DDPM sampler~\cite{DDPM_ho2020denoising}, with lower denoising steps as proposed from the denoising diffusion implicit model (DDIM)~\cite{DDIM_song2022denoising}. We use the re-scaled classifier-free-guidance formula (Sec. 3.4 of Ref.~\cite{diffFlawed_lin2023common} with $\phi=0.7$) to guide the model according to the condition. The number of denoising steps and the guidance scales used to generate the datasets presented in the figures of this manuscript are listed in Extended Data Table 3.

We note that the masking task presented in~\cref{fig:mask}a represents a bigger challenge than the rest of the tasks, due to hard restrictions imposed on the generated circuits. Indeed, we see an increase of the number of error circuits, from below $1\%$ in the non-masked entanglement generation task to $\sim 90\%$ here. Moreover, the number of samples needed to find a proper solution will vary depending on how restrictive the masking is. In contrast, the editing task presented in~\cref{fig:mask}b is not as restrictive. Hence, the number of error circuits does not increase in the case and the model is able to find multiple distinct solutions with less needed samples.

%#########################################################################
\section{Circuit length vs. SRV} \label{se:app_qc_len_srv_correlation}

Machine learning models often tend to exploit features of the training data to improve their accuracy. In this case, our experiments show that the model correlates the length of the generated circuits with the number of entangled qubits (Extended Data Fig. 4). While we balance the circuit length in the training set (see Extended Data Fig. 4a and Dataset section above), the generated circuits (Extended Data Fig. 4b and c) show peaked distributions, whose peak increases as the number of entangled qubits increase. 
As shown in the figure, this feature appears for any tensor size. However, we see that reducing the tensor size slightly improves this feature. 
Indeed, further investigations revealed that the peaks in the length distribution correspond to lengths in which randomly generated circuits would have the best accuracy. 
We can then consider that while training, the model found that producing such distribution is a shortcut to decreasing the loss function and hence biased its prediction for this.
Different dataset balancing approaches were tested, all yielding similar results.
While such preference biases the generation, the final length of the circuit can still be tailored by means of the input tensor size and the model can still produce circuits of all lengths for each SRV (if possible).

%TC:endignore
\end{document}